\documentclass[notitlepage,floats,aps,nofootinbib,preprintnumbers,twocolumn,prl,10pt,balancelastpage]{revtex4-1}

\usepackage{amsmath,amssymb,amsfonts}
\usepackage{hyperref}
\hypersetup{colorlinks,citecolor=blue}
\usepackage{graphicx}
\usepackage{xcolor}
\usepackage{subfigure}
\usepackage{nicefrac}
\usepackage{mathrsfs}
\usepackage{mdframed}
\usepackage{ulem}
\usepackage{units}
\usepackage{slashed}
\def\beq{\begin{equation}\begin{aligned}}
\def\eeq{\end{aligned}\end{equation}}

\newcommand{\Lag}{\mathcal L}
\newcommand{\Ocal}{\mathcal O}

\begin{document}

\title{Axion Portal Dark Matter and the LUX-ZEPLIN High-Recoil  Event}

\author{James Unwin}
\affiliation{Department of Physics, University of Illinois  Chicago, Chicago, IL 60607, USA}
\affiliation{Rudolf Peierls Centre for Theoretical Physics, University of Oxford, OX1 3NP, UK}

\begin{abstract}
 LUX-ZEPLIN (LZ)  has reported one event compatible with a $248~\mathrm{keV}$ nuclear recoil.
While initial attention has focused on inelastic dark matter, that
interpretation depends sensitively on the uncertain high-speed tail of the
Galactic halo.  We develop instead an elastic realisation of the
pseudoscalar-pseudoscalar interaction
$\Lag_4=(\bar\chi i\gamma^5\chi)(\bar N i\gamma^5N)$, whose contact templates
give one of the largest local significances identified by LZ.  The minimal model presented involves Dirac fermion dark matter and a pseudoscalar mediator. The pseudoscalar arises from a complex singlet and the minimal gauge-invariant UV completion also requires a single multi-TeV vector-like heavy quark.
 A representative model which realises thermal freeze-out has 
$m_\chi\sim 400~{\rm GeV}$ and $m_a\sim 1~{\rm GeV}$. The simplest model breaks from the isoscalar/isovector assumption, and the recoil spectrum lies
between LZ's elastic $\Ocal_4^v$ and contact $\Lag_4^v$ templates, suggesting
a local significance of $2.6$-$3.2\sigma$. 
\end{abstract}
\maketitle

\section{Introduction} 

\vspace{-3mm}

The LUX-ZEPLIN (LZ) Collaboration recently reported a candidate dark matter event with \cite{LZ:2026axp}
\begin{align}
 E_R&=248\pm23_{\rm stat}\pm23_{\rm sys}\ \mathrm{keV},\nonumber\\[-2pt]
 Q&\equiv|\bm q|=\sqrt{2m_{\rm Xe}E_R}\simeq246~\mathrm{MeV}.
\end{align}
Compared to a set of reference Lagrangian models \cite{Anand:2013yka}, the largest local significance is $3.4\sigma$, with  $2.6\sigma$ global significance.
Several dark matter interpretations quickly appeared~\cite{Lou:2026idn,Su:2026rwz,Yamashita:2026ump,Fan:2026kxx,Freese:2026sga,Wu:2026nhi,Yin:2026jnn,DiMauro:2026ldr,Visinelli:2026kgt,Du:2026guj,Jeesun:2026vzo,deLima:2026shq,Dent:2026bji}, with a particular focus on inelastic dark matter, especially thermal Higgsino realisations
\cite{Fan:2026kxx,Freese:2026sga,Wu:2026nhi,Yamashita:2026ump,Yin:2026jnn,DiMauro:2026ldr,Visinelli:2026kgt,Du:2026guj}.
 While elegant, the specific thermal Higgsino model appears excluded by IceCube limits on high-energy neutrinos from dark matter capture and annihilation in the Sun \cite{Pospelov:2026ewn}. More generally, inelastic dark matter remains an interesting possibility, however it is limited by the fact that the recoil rate is sensitive to the uncertainties in the high-speed tail of the Galactic halo.  This motivates elastic mechanisms in which the hard recoil spectrum arises from the interaction's momentum dependence.

Specifically, LZ tested twenty Lorentz-invariant WIMP-nucleon interactions for benchmark masses ranging from $10~{\rm GeV}$ to $4~\mathrm{TeV}$.
Of the candidate Lagrangians 
\beq
\Lag_4^i=d_4^N(\bar\chi i\gamma^5\chi)(\bar N i\gamma^5N)
\eeq is particularly attractive as it has a minimal CP-conserving
pseudoscalar completion.
For $m_\chi=200~{\rm GeV}$-$4~\mathrm{TeV}$,  the  $\Lag_4^i$ templates correspond to local significances of
$3.0$-$3.2\sigma$, with the largest due to $\Lag_4^v$ at
$m_\chi=400~{\rm GeV}$, while $\Lag_4^s$ reaches $3.1\sigma$.  Here $s$ and $v$ indicate the nucleon couplings are isoscalar or isovector.  The $\Lag_4^i$ Lagrangian term can be recast as a nonrelativistic operator via the matching  $\Lag_4 \rightarrow -d_4^N\frac{m_N}{m_\chi}\Ocal_6$ with \cite{Anand:2013yka}
\beq
 \Ocal_6&=\left(\bm S_\chi\!\cdot\!\frac{\bm q}{m_N}\right)
 \left(\bm S_N\!\cdot\!\frac{\bm q}{m_N}\right).
\eeq
In contrast, most of the inelastic formulations, and in particular the Higgsino models, map to $\mathcal{O}_1$ with mass splitting $\delta \sim 300$ keV.

In what follows, we present the first full model that UV completes one of the elastic operators favoured by LZ. We show that the $\Lag_4$ interaction naturally maps to the axion portal \cite{Nomura:2008ru}, and that this realisation can simultaneously reproduce both the LZ event and the observed dark matter abundance via thermal freeze-out. Additionally, we highlight complementary observables for this scenario.

\vspace{-2mm}
\section{pNGB realisation of $\mathcal L_4$}
\vspace{-2mm}

We supplement the Standard Model with a gauge singlet complex scalar
$ \Phi=(1/\sqrt2)(f+\rho)e^{ia/f}$
alongside a Dirac fermion $\chi$ with the interaction Lagrangian
\beq
 \Lag_\chi=-y_\chi\Phi\bar\chi_L\chi_R+\mathrm{h.c.}
\eeq
A U(1)$_X$ forbids a bare Dirac mass, which is broken by the VEV of $\Phi$.  After symmetry breaking, $ m_\chi=y_\chi f/\sqrt2$ and $g_\chi\equiv m_\chi/f$ with
\beq
 \Lag_\chi=-m_\chi\left(1+\frac{\rho}{f}\right)
 \bar\chi e^{i\gamma^5a/f}\chi .
\eeq
We identify $a$ as the pseudoscalar mediator and take
$m_a<m_\chi$ and $m_\rho+m_a>2m_\chi$. Thus
$\chi\bar\chi\to aa$ is open while the on-shell $a\rho$ channel is closed,
although off-shell $\rho$ exchange still contributes to freeze-out.
Before explicit symmetry breaking, $a$ is the massless Goldstone boson of $U(1)_X$.
The potential for $a$  is generated
only by terms that explicitly break $U(1)_X$. At tree level, we introduce the real soft-breaking term 
\begin{equation}
 V_{\rm br}=-\mu^3(\Phi+\Phi^\dagger),
 \label{eq:softmass}
\end{equation}
leading to the bare contribution
$m_{a,0}^2=\sqrt2\mu^3/f$.  Below we denote the physical mass, including
radiative corrections, by $m_a^2=m_{a,0}^2+\delta m_a^2$.
 Equation~\eqref{eq:softmass} selects
a unique CP-preserving vacuum, so it produces neither spontaneous CP violation nor
stable domain walls. 

For real $\mu^3>0$, the explicit-breaking potential is
\beq
 V_{\rm br}=-\sqrt2~\mu^3(f+\rho)\cos(a/f),
\eeq
and has a single physical minimum at $\langle a\rangle=0$ up to periodicity
$a\rightarrow a+2\pi f$.  Moreover, both $a$ and
$\bar\psi i\gamma^5\psi$ are CP odd, so the products
$a\bar\chi i\gamma^5\chi$ and $a\bar u i\gamma^5u$ are CP even.  
The new sector therefore introduces no independent CP-violating phase and generates no new electric dipole moments.

We next specify the couplings of the pseudoscalar mediator to the Standard Model. 
An isoscalar coupling, with $g_u=g_d$, is attractive since it has a similar recoil spectrum to the contact operator (as we discuss in the next section).
However, reproducing the LZ event in the isoscalar case requires large couplings
and for $2~\mathrm{TeV}$ messengers, their explicit symmetry-breaking couplings generate 
$|\delta m_a|\sim20$-$25~\mathrm{GeV}$, so a $1~\mathrm{GeV}$ mediator
requires a 1\% tuning between $m_{a,0}^2$ and $\delta m_a^2$. Moreover, $g_u=g_d$
must hold at the few-percent level to prevent the pion pole from reappearing.
We therefore focus on the $u$-only portal: it requires only one vector-like
messenger and preserves the radiative motivation for a GeV-scale pseudoscalar.

For simplicity, we assume that the only coupling is via the up-quark bilinear
\begin{equation}
 \Lag_a^u= g_u a\bar u i\gamma^5u.
\end{equation}
The above interaction holds below the scale of electroweak symmetry breaking and arises from the gauge-invariant dimension-five operator $ia~\bar q_{1L}\widetilde H u_R+\mathrm{h.c.}$. Such a dimension five operator can be generated by integrating out a vector-like Dirac $U$ messenger.
This structure can be enforced by an approximate $Z_2^u$ symmetry under
which $U_{L,R}$, $q_{1L}$, and $u_R$ are odd, while all other fields are
even.  The down-quark Yukawa and CKM mixing then arise from small
$Z_2^u$-breaking spurions, which induce correspondingly suppressed
couplings to other quarks.

We introduce one vector-like Dirac up-type messenger,
$U=(U_L,U_R)$, with $U_{L,R}\sim(\bm3,\bm1,2/3)$.
The CP-invariant messenger Lagrangian is
\begin{equation}
 \Lag_U=-M_U\bar UU-y_L\bar q_{1L}\widetilde H U_R
 -iy_RA\bar U_Lu_R+\mathrm{h.c.}
\end{equation}
where
$ A\equiv\frac{\Phi-\Phi^\dagger}{i\sqrt2}
 =(f+\rho)\sin(a/f)$.
Integrating out $U$ gives the gauge-invariant operator
\beq
 \Lag_{\rm eff}=i\frac{y_Ly_R}{M_U}A
 (\bar q_{1L}\widetilde Hu_R)+\mathrm{h.c.}
\eeq
with $g_u=\frac{y_Ly_Rv}{\sqrt2M_U}$.
The $y_R$ interaction breaks $U(1)_X$ and generates a mass correction
\begin{equation}
 |\delta m_a|\sim
 \left(\frac{N_cy_R^2M_U^2}{8\pi^2}\right)^{1/2}
 \simeq1~{\rm GeV}
 \left(\frac{y_R}{3\times10^{-3}}\right)
 \left(\frac{M_U}{2{\rm TeV}}\right).
 \label{eq:visiblemass}
\end{equation}
The interaction responsible for the LZ signal also contributes to the
mediator mass.  If the mediator is made heavier, a stronger coupling is
needed to maintain the observed scattering rate, which in turn generates an
even larger mass correction.  Avoiding this feedback naturally favours
$m_a\lesssim2~{\rm GeV}$.  Moreover, for $m_a<3m_{\pi^0}\simeq0.405~{\rm GeV}$ the mediator can become long-lived, and
beam-dump and cosmological considerations are constraining.  Thus we focus on the
natural and phenomenologically safer range $m_a\sim0.5$-$2~{\rm GeV}$.

The generic renormalisable portal $\lambda_{\Phi H}|\Phi|^2|H|^2$ mixes the radial mode with
the Higgs, with 
\beq
\theta_{h\rho}\simeq\lambda_{\Phi H}fv/(m_\rho^2-m_h^2).
\eeq 
For $f\simeq513~{\rm GeV}$ and $m_\rho=1~{\rm TeV}$, this gives
$\theta_{h\rho}\simeq0.13\lambda_{\Phi H}$.
This leads to a coherent scattering cross-section of order 
\beq
\sigma_{\rm SI}\simeq6\times10^{-47}~{\rm cm}^2~\left(\frac{\lambda_{\Phi H}}{0.1}\right)^2\left(\frac{1~{\rm TeV}}{m_\rho}\right)^4.
\eeq
Existing spin-independent limits \cite{LZ:2024zvo} at $m_\chi=400$ GeV require $|\lambda_{\Phi H}|\lesssim0.07(m_\rho/1~{\rm TeV})^2$.  To avoid exclusion, we require that this mixed quartic be suitably small; notably, couplings of $\mathcal{O}(0.01)$ are commonplace in the Standard Model. 
Moreover, the $\rho$-mediated scattering may provide a secondary discovery channel in next-generation direct detection experiments, such as XLZD \cite{XLZD:2024nsu}.

While for inelastic dark matter \cite{Su:2026rwz,Yamashita:2026ump,Fan:2026kxx,Freese:2026sga,Wu:2026nhi,Yin:2026jnn,DiMauro:2026ldr,Visinelli:2026kgt,Du:2026guj,deLima:2026shq,Dent:2026bji} the hardness in the recoil spectrum is due to inelastic kinematics, in the models studied here, the hardness arises from momentum dependence (potentially moderated by the pion pole).


\vspace{-2mm}
\section{Recoil spectrum}
\vspace{-2mm}

The momentum dependence of the quark-nucleon matching is described by
\begin{equation}
 \langle N|\bar u i\gamma^5u|N\rangle
 =\bar g_u^N(Q^2)~\bar N i\gamma^5N .
\end{equation}
At the momentum of the LZ event, $Q_0=246~\mathrm{MeV}$, representative
matching gives \cite{Bishara:2017pfq}:
$ \bar g_u^{p,n}(Q_0^2)\simeq(+44,-41)$.
Exchange of the mediator then gives
\begin{equation}
 d_4^N(Q^2)=
 \frac{g_\chi g_u~\bar g_u^N(Q^2)}{m_a^2+Q^2}.
\end{equation}
Using the public xenon $\Sigma''$ nuclear responses implemented in
\textsc{WIMpy\_NREFT}~\cite{WIMpy}, an indicative one-event
normalisation requires
\beq
 d_{4,\mathrm{req}}^{(42)}(Q_0^2)
 \simeq(2\text{-}4)\times10^{-3}~\mathrm{GeV}^{-2},
\eeq
where the superscript denotes $|\bar g_u^N|=42$.  Equivalently,
\begin{align}
 g_\chi g_u\simeq{}&8\times10^{-5}
 \left(\frac{d_{4,\mathrm{req}}^{(42)}}
 {3\times10^{-3}~\mathrm{GeV}^{-2}}\right)
 \left(\frac{m_a^2+Q_0^2}{1.06~\mathrm{GeV}^2}\right).
 \label{eq:LZnorm}
\end{align}
This estimate carries nuclear-response and detector-normalisation
uncertainties and is not a replacement for the LZ likelihood.

\begin{figure}[t]
 \centering
 \includegraphics[width=0.95\columnwidth]{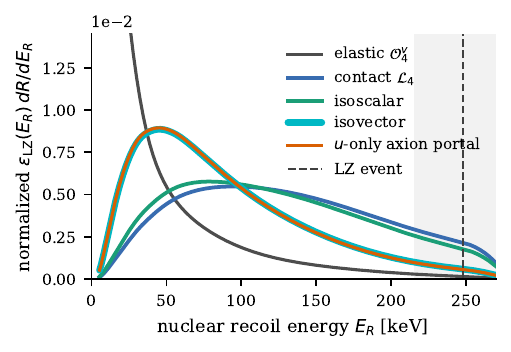}
\vspace{-3mm} \caption{Normalized recoil spectra relevant for the LZ event. Model parameters are fixed as in Eq.~(\ref{benchmark}), in particular, the dark matter mass is $400~\mathrm{GeV}$ and the pseudoscalar mediator mass is $1$ GeV.
 The published nuclear-recoil efficiency is applied, but the full detector response is not included.}
\vspace{-2mm}
 \label{fig:spectra}
\end{figure}

The up-quark pseudoscalar density contains an isovector component that can
create a virtual neutral pion.  The pion propagator therefore contributes
$(Q^2+m_{\pi^0}^2)^{-1}$ to the scattering amplitude
\cite{Bishara:2017pfq}.  Keeping the leading one-body pole contribution gives
\begin{equation}
 \frac{dR}{dE_R}\propto
 \frac{Q^4}
 {(Q^2+m_{\pi}^2)^2(m_a^2+Q^2)^2}
 W_{\Sigma''}(Q^2)~\eta(v_{\min}),
\end{equation}
where $m_\pi\approx134.98~\mathrm{MeV}$.  The explicit $Q^4$ ensures that the
rate still vanishes at threshold.  For $Q\gtrsim m_\pi$, however, the pion
propagator compensates this rise, so the spectrum saturates and subsequently
falls with the halo integral and nuclear response.

Figure~\ref{fig:spectra} shows the recoil spectra evaluated over the LZ analysis window, $5.4<E_R<269.9~{\rm keV}$, using
\textsc{WIMpy\_NREFT}~\cite{WIMpy}.  This package supplies the public
$^{129}{\rm Xe}$ and $^{131}{\rm Xe}$ response functions,
as well as the recoil kinematics, Standard Halo Model velocity
integral, and conversion of proton and neutron NREFT coefficients into
$dR/dE_R$~\cite{Fitzpatrick:2012ix}. 
 Each spectrum is then multiplied by an
interpolation of the NR efficiency published in Fig.~S2 of
Ref.~\cite{LZ:2026axp}. 
Figure~\ref{fig:spectra} specifically shows the resulting unit-normalized spectrum for
$m_\chi=400~\mathrm{GeV}$, together with the elastic $\Ocal_4^v$ and contact
$\Lag_4^v$ templates.  The pion-pole spectrum contains
fewer low-energy recoils than $\Ocal_4^v$, but is softer than contact
$\Lag_4^v$.  Each curve has unit integral over the displayed range, so its
height represents the conditional recoil-energy distribution rather than an
absolute event rate.

Observe that the $u$-only model has a similar structure to the isovector due to the presence of the pion pole, while the isoscalar more closely matches the contact operator (but requires $\sim1$\% finetuning in $m_a$ to realise.) Moreover, the models considered here are sandwiched by two benchmark templates of the LZ study  \cite{LZ:2026axp}.   The LZ analysis found a local significance of $2.6\sigma$ for elastic $\Ocal_4^v$ (with $\delta=0$) and $3.2\sigma$ for the contact operator
$\Lag_4^v$, this suggests the template-level bracket for the local significance of the $u$-only and isovector models is
\beq
 2.6\lesssim Z_{\mathrm{loc}}\lesssim3.2 .
\eeq
 A precise significance statement would require a full likelihood analysis and is beyond the scope of this work.

\vspace{-2mm}
\section{Thermal freeze-out}
\vspace{-2mm}

For  $m_a<m_\chi$ and $m_\rho+m_a>2m_\chi$ the dominant freeze-out process is $ \chi\bar\chi\to aa$.
Annihilation into Standard Model quarks is negligible because of the small coupling $g_u$.  
Expanding the nonlinear mass term gives
\begin{equation}
 \Lag_\chi\supset
 -g_\chi a\bar\chi i\gamma^5\chi
 +\frac{g_\chi^2}{2m_\chi}a^2\bar\chi\chi+\cdots .
 \label{eq:interactions}
\end{equation}
The annihilation amplitude receives contributions from $t$- and $u$-channel
$\chi$ exchange together with the four-point interaction in
Eq.~\eqref{eq:interactions}; all three contributions must be
retained to respect the nonlinear symmetry.  
In the decoupled-$\rho$ limit the leading annihilation channel
is $p$-wave, $ \sigma v_{\rm rel}
 =b_\infty(r)v_{\rm rel}^2+\mathcal O(v_{\rm rel}^4)$, with $r=m_a/m_\chi$ and 
\beq
 b_\infty(r)=\frac{g_\chi^4}{384\pi m_\chi^2}\sqrt{1-r^2}
 \left[2+\frac{r^8}{(2-r^2)^4}\right],
 \label{eq:annihilation}
\eeq
This includes $t/u$-channel $\chi$ exchange and the $a^2\bar\chi\chi$ four-point interaction.
 In the limit $r\ll1$ this reduces to
\beq
b_\infty(0)=\frac{g_\chi^4}{192\pi m_\chi^2}.
\eeq

Even while on-shell  $\chi\bar\chi\to a\rho$ is closed, the $\rho$ mode does not entirely decouple from freeze-out.  Off-shell $s$-channel $\rho$ exchange combines with the four-point
interaction of Eq.~\eqref{eq:interactions}, modifying the matrix element to
\begin{equation}
 {\cal M}_{\rho}=
 \frac{g_\chi^2}{m_\chi}\bar vu
 \left[1-\frac{s-2m_a^2}
 {s-m_\rho^2+i m_\rho\Gamma_\rho}\right].
\end{equation}
Accordingly, for $m_a\ll m_\chi$, the $p$-wave coefficient is
enhanced relative to the decoupled-$\rho$ case $b_\infty$ by a factor
\begin{equation}
 \mathcal{F}_\rho=\frac32 |C_\rho|^2-\rm{Re}[C_\rho]+\frac12,
\end{equation}
with 
\beq
C_\rho=
1-\frac{4m_\chi^2-2m_a^2}
{4m_\chi^2-m_\rho^2+i m_\rho\Gamma_\rho},
\eeq 
In the limit $m_a\ll m_\chi$, one has $ C_\rho=m_\rho^2/(m_\rho^2-4m_\chi^2)$. 
Taking $m_\chi\simeq400~{\rm GeV}$ and $m_\rho\simeq1~{\rm TeV}$ gives
$\mathcal F_\rho\simeq9.3$.

\begin{figure}[t]
 \centering
 \includegraphics[width=0.97\columnwidth]{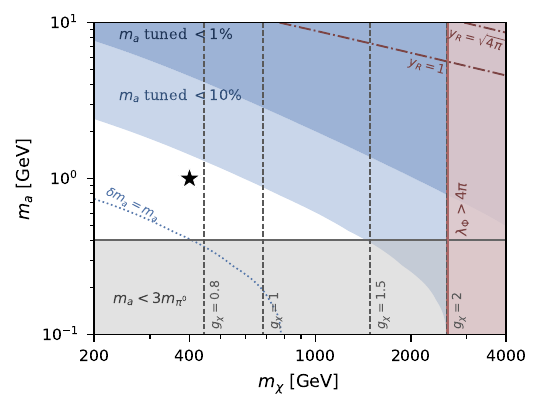}
\vspace{-6mm} \caption{Parameter space of the $u$-only axion portal in the $(m_\chi,m_a)$ plane. At each point, the observed relic abundance fixes $g_\chi$ through
$p$-wave $\chi\bar\chi\to aa$ freeze-out, using the threshold enhancement $\mathcal F_\rho=9.3$, corresponding to
$m_\rho=2.5m_\chi$. The LZ one-event normalisation fixes $g_u$, and taking $y_L=0.2$ and $M_U=2$ TeV we show the corresponding messenger coupling  $y_R$. The light (dark) blue shading indicates where an appropriate $m_a$ requires a sub-$10\%$ (sub-$1\%$) cancellation between the bare contribution and the radiative correction of Eq.~\eqref{eq:visiblemass}. In the grey region $m_a<3m_{\pi^0}$ and the leading three-pion channel closes, implying beam-dump and cosmological constraints must be considered. Taking $m_\rho=2.5m_\chi$, in the red region the radial quartic $\lambda_\Phi=m_\rho^2/2f^2$ exceeds $4\pi$. In the white region the model accounts for both the LZ event and the dark matter abundance with  $m_a$ untuned. The star ($\star$) marks the benchmark model outlined in Eq.~\eqref{benchmark}. }
\vspace{-2mm}
 \label{fig:parameter-space}
\end{figure}

For a symmetric Dirac population, with
$n=n_\chi+n_{\bar\chi}$, the Boltzmann equation is
\begin{equation}
 \dot n+3Hn
 =-\frac{1}{2}\langle\sigma v_{\rm rel}\rangle
 \left(n^2-n_{\rm eq}^2\right).
 \label{eq:boltzmann}
\end{equation}
 Since
$\langle v_{\rm rel}^2\rangle\simeq6/x$, with
$x=m_\chi/T$, the thermally averaged cross-section is
$\langle\sigma v_{\rm rel}\rangle\simeq6b/x$ with
$b=\mathcal F_\rho\,b_\infty(0)$.
Using the standard freeze-out approximation to 
Eq.~\eqref{eq:boltzmann}, with $x_f\simeq26$ and
$g_*\simeq90$, and fixing $\Omega_\chi h^2\simeq0.12$ gives
\beq
g_\chi &\simeq0.8
\left(\frac{m_\chi}{400\,{\rm GeV}}\right)^{1/2}
\left(\frac{x_f}{26}\right)^{1/2}
\left(\frac{9.3}{\mathcal F_\rho}\right)^{1/4}\\
f&\simeq513\,{\rm GeV}
\left(\frac{m_\chi}{400\,{\rm GeV}}\right)^{1/2}
\left(\frac{26}{x_f}\right)^{1/2}
\left(\frac{\mathcal F_\rho}{9.3}\right)^{1/4}.
\label{eq:DM}
\eeq
Figure~\ref{fig:parameter-space} illustrates the restrictions on the parameter space. The white region indicates where the model accounts for both the dark matter abundance and the LZ event with  $m_a$ untuned.
Notably, the dominant \(aa\) channel is \(p\)-wave suppressed at late times and the residual \(s\)-wave annihilation into \(u\bar u\) is negligible for \(g_u\sim10^{-4}\).
For the central normalisation in Eq.~\eqref{eq:LZnorm}, a representative point which realises the full model
 with $g_\chi \simeq0.78$ and $f\simeq513\,{\rm GeV}$ as indicated in Eq.~\eqref{eq:DM} is given by
\beq
 m_\chi &=400~{\rm GeV}, \hspace{3mm}
 m_a=1~{\rm GeV},   \hspace{7mm}  m_\rho = 1~{\rm TeV}, \\
 M_U&=2~{\rm TeV}, \hspace{8mm} g_u=9.7\times10^{-5}, \hspace{3mm}   \lambda_\Phi = 1.9,\\
y_L &= 0.2,  \hspace{14mm} y_R = 5.6\times10^{-3}.
 \label{benchmark}
\eeq
QCD pair production gives the leading collider test on the heavy coloured mediators.
This benchmark model lies above the $1.53~{\rm TeV}$ Run-2 limit \cite{ATLAS:2024zlo}.

\vspace{-3mm}
\section{Concluding remarks}
\vspace{-2mm}

While the LZ observation is only a single event with a $2.6\sigma$ global significance and
may still be background, there remains the prospect that it could be the first laboratory signal of dark matter.
Taking the one-event best-fit normalisation, Poisson statistics imply a 95\% probability of at least one additional signal event in the analysis window after $8.5\,{\rm tonne\,yr}$ ($660$ days) of further exposure.

 If further events are observed, the pseudoscalar mediator proposed here provides an economical
elastic interpretation.  
In contrast to the inelastic interpretations emphasized in the current literature, the high-recoil spectrum arises from momentum-dependent elastic scattering, rather than inelastic kinematics \cite{Su:2026rwz,Yamashita:2026ump,Fan:2026kxx,Freese:2026sga,Wu:2026nhi,Yin:2026jnn,DiMauro:2026ldr,Visinelli:2026kgt,Du:2026guj,Jeesun:2026vzo,deLima:2026shq,Dent:2026bji}.
The pion pole inherent to the isovector and $u$-only models makes the recoil
spectrum softer than contact $\Lag_4$ but harder than elastic $\Ocal_4$.
The LZ template models  \cite{LZ:2026axp} thus characteristically bound the local significance:
$2.6\lesssim Z_{\rm loc}\lesssim3.2$; a full likelihood analysis is needed for a precise value.

We stress that the `axion' appearing here is not identified with the QCD axion and does not address the strong-CP problem. Rather, it is an axion-like particle (ALP), the pNGB of a spontaneously broken global symmetry. Such states are ubiquitous in extensions of the Standard Model, with their masses controlled by explicit symmetry breaking. The axiverse proposal \cite{Arvanitaki:2009fg} suggests that $\mathcal{O}$(100) axions could arise in string theory. Moreover, Nomura \& Thaler \cite{Nomura:2008ru} also suggest in passing an $R$-axion portal (the pNGB of a spontaneously broken global U(1)${}_R$ symmetry within supersymmetry \cite{Bagger:1994hh}), which can be potentially light with distinct phenomenology \cite{Bellazzini:2017neg,Unwin:2026nqn}.

Interestingly, the vector-like quark required in the model leads to a correlated modification of charged-current observables. Mixing between $U$ and $u$ occurs with $s_L\simeq y_Lv/(\sqrt{2}M_U)$, suppressing the first row of the observed CKM matrix and giving $\Delta_{\rm CKM}\equiv\sum_j|V_{uj}|^2-1\simeq-s_L^2$. Intriguingly, the benchmark above implies a shift $\Delta_{\rm CKM}\simeq-3\times10^{-4}$, which has the same sign as the reported $\sim3\sigma$ first-row CKM unitarity deficit \cite{Crivellin:2022rhw}. However, a full study of this `Cabibbo-angle anomaly' and the present model is left for future work.

 Different dark matter scenarios can be disambiguated via their dependence on nuclear spin, isospin and the pion-pole form factor, and thus by comparing xenon with a different spin-sensitive target, such as tungsten in a dedicated high-recoil CRESST analysis \cite{Angloher:2025fzw}.
Moreover, future experiments, such as XLZD \cite{XLZD:2024nsu}, will provide further discrimination. Additionally, the UV completion outlined here requires $\sim2$ TeV coloured states, which are potentially observable at near-future colliders and can provide a complementary~probe.

\vspace{1mm}\noindent {\bf Acknowledgements.}~The author is grateful to John Wheater for helpful interactions, also to New College, Queen's College, and the Rudolf Peierls Centre for their hospitality. Claude \& Codex aided in the analysis.

\end{document}